\documentclass[12pt]{article}   
\pdfoutput=1
\usepackage{geometry,bbm}   
\usepackage{ulem}
\usepackage[toc,page]{appendix}             		
\usepackage{graphicx,cancel}
\usepackage[usenames,dvipsnames]{color}
\usepackage{slashed}					
\usepackage{amssymb,amsmath}
\usepackage{hyperref}
\usepackage{mathtools}
\usepackage{cite}

\usepackage{graphicx}
\usepackage{bm}
\def\bea{\begin{eqnarray}}
\def\eea{\end{eqnarray}}
\usepackage{latexsym}
\usepackage{epsfig,amssymb,euscript}
\usepackage{amsmath}
\usepackage{array,calc,epsfig}
\usepackage{bbold}

\newcommand{\be}{\begin{equation}}
\newcommand{\ee}{\end{equation}}

\numberwithin{equation}{section}

\begin{document}

\begin{titlepage}

\begin{center}
{\LARGE
{\bf
Supersymmetry breaking deformations and \vskip 20pt phase transitions in five dimensions
}}
\end{center}

\bigskip
\begin{center}
{\large
Matteo Bertolini and Francesco Mignosa}
\end{center}

\renewcommand{\thefootnote}{\arabic{footnote}}

\begin{center}
\vspace{0.2cm}
SISSA \\
Via Bonomea 265, I 34136 Trieste, Italy\\
and
\\
INFN, Sezione di Trieste \\ Via Valerio 2, I-34127 Trieste, Italy

\vskip 5pt
{\texttt{bertmat@sissa.it, fmignosa@sissa.it}}
\end{center}

\vskip 5pt
\noindent
\begin{center} {\bf Abstract} \end{center}
\noindent
We analyze a recently proposed supersymmetry breaking mass deformation of the $E_1$ superconformal fixed point in five dimensions which, at weak gauge coupling, leads to pure $SU(2)$ Yang-Mills and which was conjectured to lead to an interacting CFT at strong coupling. We provide an explicit geometric construction of the deformation using brane-web techniques and show that for large enough gauge coupling a global symmetry is spontaneously broken and the theory enters a new phase which, at infinite coupling, displays an instability. The Yang-Mills and the symmetry broken phases are separated by a phase transition. Depending on the structure of the potential, this can be first or second order. 

\vspace{1.6 cm}
\vfill

\end{titlepage}

\newpage
\tableofcontents

\section{Introduction and summary of results}
Recently there has been a renewed interest in the study of the dynamics of five-dimensional (supersymmetric) field theories pioneered in \cite{Seiberg:1996bd,Morrison:1996xf,Intriligator:1997pq,Aharony:1997ju,Aharony:1997bh,DeWolfe:1999hj}, especially as far as strongly coupled supersymmetric fixed points are concerned, see e.g. \cite{Cremonesi:2015lsa,Bergman:2015dpa,Zafrir:2015ftn,Gutperle:2017nwo,Gutperle:2018axv,Hayashi:2018bkd,Hayashi:2018lyv,Hayashi:2019yxj,Xie:2017pfl,Jefferson:2018irk,Bergman:2018hin,Closset:2018bjz,Bhardwaj:2018yhy,Bhardwaj:2018vuu,Bhardwaj:2019jtr,Bhardwaj:2019xeg,Bhardwaj:2020gyu,Closset:2020scj,Bhardwaj:2020ruf,Bhardwaj:2020avz,Apruzzi:2019vpe,Apruzzi:2019opn,Apruzzi:2019enx,Apruzzi:2019kgb,Closset:2019juk,vanBeest:2020kou,vanBeest:2020civ,Bergman:2020myx,Closset:2020afy}. A still open question is whether non-supersymmetric ones exist. Investigations in this direction have been done mostly using the $\epsilon$-expansion approach and bootstrap techniques \cite{Fei:2014yja,Nakayama:2014yia,Bae:2014hia,Chester:2014gqa,Li:2016wdp,Arias-Tamargo:2020fow,Li:2020bnb,Giombi:2019upv,Giombi:2020enj}, but it is fair to say that we are still far from having any clear picture about the existence of interacting conformal field theories (CFT) in five dimensions.  

A concrete route one can explore is to start from known superconformal field theories (SCFT) and softly break supersymmetry. In principle this could lead, at least in certain regions of the parameter space, to non-supersymmetric interacting fixed points. 

In \cite{BenettiGenolini:2019zth} a supersymmetry breaking mass deformation of the $E_1$ theory was considered. The $E_1$ theory is a SCFT which can be seen as a UV completion of five-dimensional $SU(2)$ SYM \cite{Seiberg:1996bd}. More precisely, upon a supersymmetric mass deformation by parameter $h=1/g^2$, the $E_1$ theory flows to pure $SU(2)$ SYM with gauge coupling $g$. The authors of \cite{BenettiGenolini:2019zth} considered a supersymmetry breaking deformation of the $E_1$ theory parameterized by a mass (squared) parameter $\widetilde m$. 

\begin{figure}[h!]
	\centering
	\includegraphics[scale=0.38]{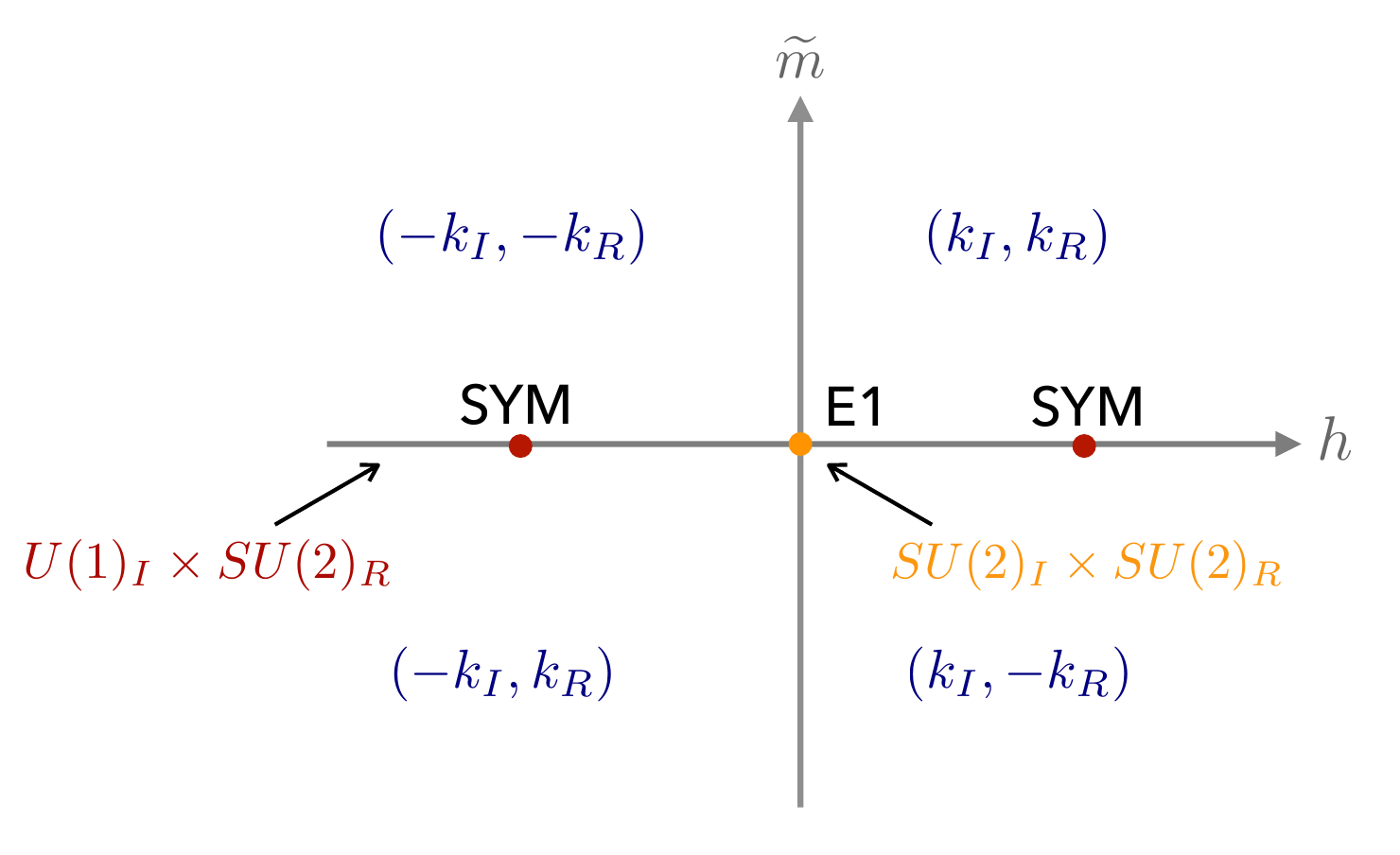}
	\caption{Global CS levels of the mass deformed $E_1$ theory. At generic points on the two-dimensional plane there is a $U(1)_I \times U(1)_R$ global symmetry, which is enhanced to $U(1)_I \times SU(2)_R$ on the $h$-axis and to $SU(2)_I \times SU(2)_R$ at the origin, where the superconformal phase is realized. For $\widetilde m=0$ one recovers ${\cal N}=1$ supersymmetry. When crossing the axes, the CS levels $(k_I,k_R)$ jump as indicated. 
	}
	\label{ps_0}
\end{figure}

While at weak coupling the supersymmetry breaking deformation leads to pure $SU(2)$ YM, it was argued that at strong coupling the theory may flow to an interacting CFT. This was suggested from the study of some topological properties of the mass-deformed $E_1$ theory in the plane $(h,\widetilde m)$. For  generic values of $h$ and $\widetilde m$ the theory enjoys a $U(1)_I \times U(1)_R$ global symmetry, a topological symmetry and a flavor symmetry, respectively.\footnote{The subindex $R$ is to recall that this abelian factor originates from the $SU(2)$ R-symmetry of the supersymmetric theory, which is explicitly broken by the supersymmetry breaking deformation.} By adding background gauge fields for these abelian symmetries one can compute the Chern-Simons (CS) levels of the corresponding background gauge fields and find that they jump across the axes, following a pattern as described in figure \ref{ps_0}. 

The jump across the $h$-axis affects only the $U(1)_R$ CS level and is understood perturbatively, in terms of gaugini becoming massless on the $h$-axis (where supersymmetry is restored). On the contrary, no such explanation holds in crossing the $\widetilde m$-axis, where the jump in the CS levels occurs also for the topological $U(1)_I$ symmetry, under which perturbative states are neutral. 
As argued in \cite{BenettiGenolini:2019zth}, there are basically three alternative scenarios which may account for such jump:
\begin{itemize}
\item The $U(1)_I \times U(1)_R$ global symmetry is preserved on the $\widetilde m$-axis and a second order phase transition holds there, with massless matter charged under $U(1)_I \times U(1)_R$. 
\item A discrete $\mathbb{Z}_2$ symmetry acting on the gauge coupling $h$ as $h\rightarrow -h$ is spontaneously broken on the $\widetilde m$-axis, indicating a first order phase transition where two iso-energetic vacua carrying different CS levels coexist.  
\item The $U(1)_I \times U(1)_R$ global symmetry is spontaneously broken on the $\widetilde m$-axis and the system enters a non-linear-sigma-model (nl$\sigma$m) phase.
\end{itemize}

In this paper, using field theory and 5-brane web techniques \cite{Aharony:1997ju,Aharony:1997bh}, we give evidence that the third scenario is the one that is actually realized. On the $\widetilde m$ axis, or more precisely in a region symmetrically displaced around it, some otherwise massive scalar condenses and the global $U(1)_I \times U(1)_R$ symmetry is broken to a diagonal $U(1)$. We also show that at infinite coupling, namely for $h=0$, there is an instability, the scalar field VEV is pushed all the way to infinity and the potential is unbounded from below. At finite $h$, instead, the minimum of the potential may be stabilized and the scalar field VEV be at finite distance in field space. 

The symmetry broken phase is separated from the pure YM phase by a phase transition. This can be first or second order, depending on the exact expression of the potential, which however we cannot determine.

Our aim in the rest of the paper is to provide evidence for the validity of this picture. Section \ref{E1sum} contains a recap of 5d $SU(2)$ SYM, of its UV-completion and of its moduli space, which includes both a Coulomb branch and a Higgs branch, which opens-up at strong coupling. In doing so, we will also offer some improvements of the 5-brane web description of the $E_1$ theory and its moduli space. For example, we will characterize the hypermultiplets describing the Higgs branch. These will play a crucial role for what comes in section \ref{susybr}, where we discuss the supersymmetry breaking mass deformation proposed in \cite{BenettiGenolini:2019zth}, its description in terms of 5-brane web and its effects on the IR dynamics. Finally, in section \ref{phdiag} the resulting phase diagram as a function of $h$ and $\widetilde m$ is presented, together with an outlook. 

\section{The $\mathbf{E_1}$ theory and SU(2) SYM}
\label{E1sum}

In \cite{Seiberg:1996bd} it was argued that ${\cal N}=1$ $SU(2)$ pure SYM, which is a non-renormalizable, IR-free theory in five dimensions, admits a UV completion in terms of  an interacting SCFT, known as  $E_1$ theory. The $E_1$ theory admits a moduli space made of a real one-dimensional Coulomb branch and a quaternionic one-dimensional Higgs branch, as well as supersymmetric and non-supersymmetric mass deformations.  

In this section we review different aspects of the $E_1$ theory using 5-branes webs. Following \cite{Aharony:1997ju,Aharony:1997bh}, whose conventions we adopt,  we first recall the structure of the brane web which realizes this theory and its supersymmetric mass deformation which is pure $SU(2)$ SYM, emphasizing the uselfullness of adding 7-branes to the brane web, as originally discussed in \cite{DeWolfe:1999hj}. Then, we will  describe the moduli space of the $E_1$ theory, also offering some improvements of its brane web description known so far. 

\begin{figure}[h!]
	\centering
	\includegraphics[scale=0.28]{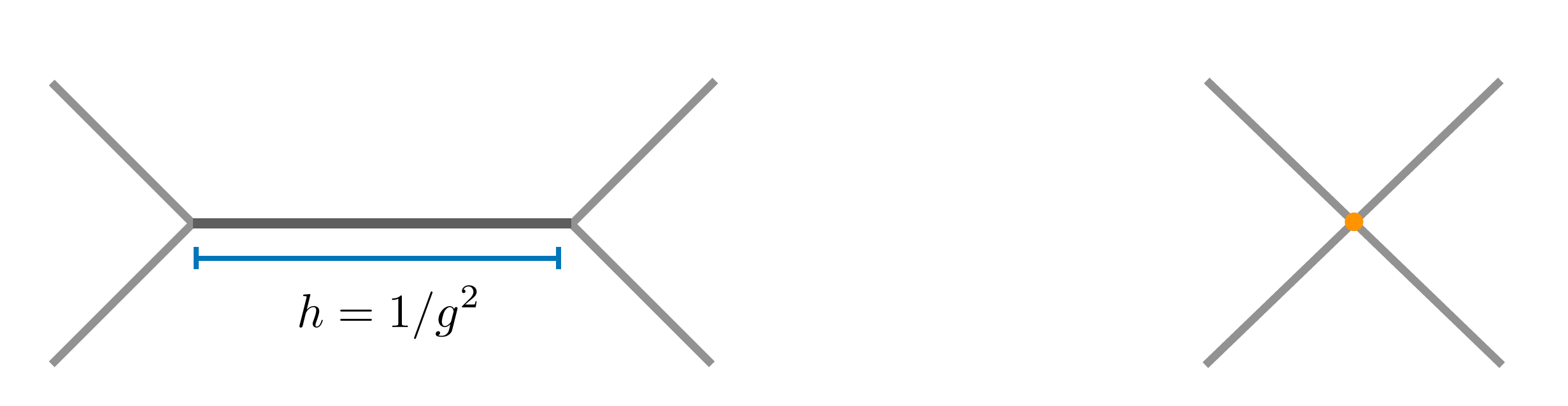}
	\caption{On the left the 5-brane web description of ${\cal N}=1$ $SU(2)$ pure SYM (the horizontal segment represents two D5-branes on which the $SU(2)$ gauge theory lives). The finite length of the D5 on the $(x,y)$ plane is proportional to the inverse gauge coupling squared. On the right the strongly coupled $E_1$ fixed point. The external branes are all semi-infinite. From the $E_1$ theory point of view $h$ acts a supersymmetric mass deformation.}
	\label{pq_1}
\end{figure}

Both ${\cal N}=1$ $SU(2)$ pure SYM and its UV fixed point can be described in terms of 5-brane webs, as depicted in figure \ref{pq_1}. All 5-branes in the figure share five space-time dimensions along $(01234)$ and meet at points (brane junctions) on a two-dimensional plane, which we parametrize by $(x,y)$. Hence a 5-brane appears as a segment in the $(x,y)$ plane with an angle which depends on its (p,q) charges (for instance, a D5-brane, namely a (1,0) 5-brane, is an horizontal segment while a NS5-brane, a (0,1) 5-brane, a vertical one). There are then three remaining Dirichlet directions along $(789)$ common to all branes. Rotations in such three-dimensional space correspond to the global $SU(2)_R$ R-symmetry. There exists another global (instantonic) symmetry. It is a $U(1)_I$ symmetry which gets enhanced to $SU(2)_I$ at the $E_1$ interacting fixed point. This symmetry and its enhancement are not visible in this description. To make the instantonic symmetry visible, the brane web must be enriched adding 7-branes, as we now discuss.

\begin{figure}[h!]
	\centering
	\includegraphics[scale=0.27]{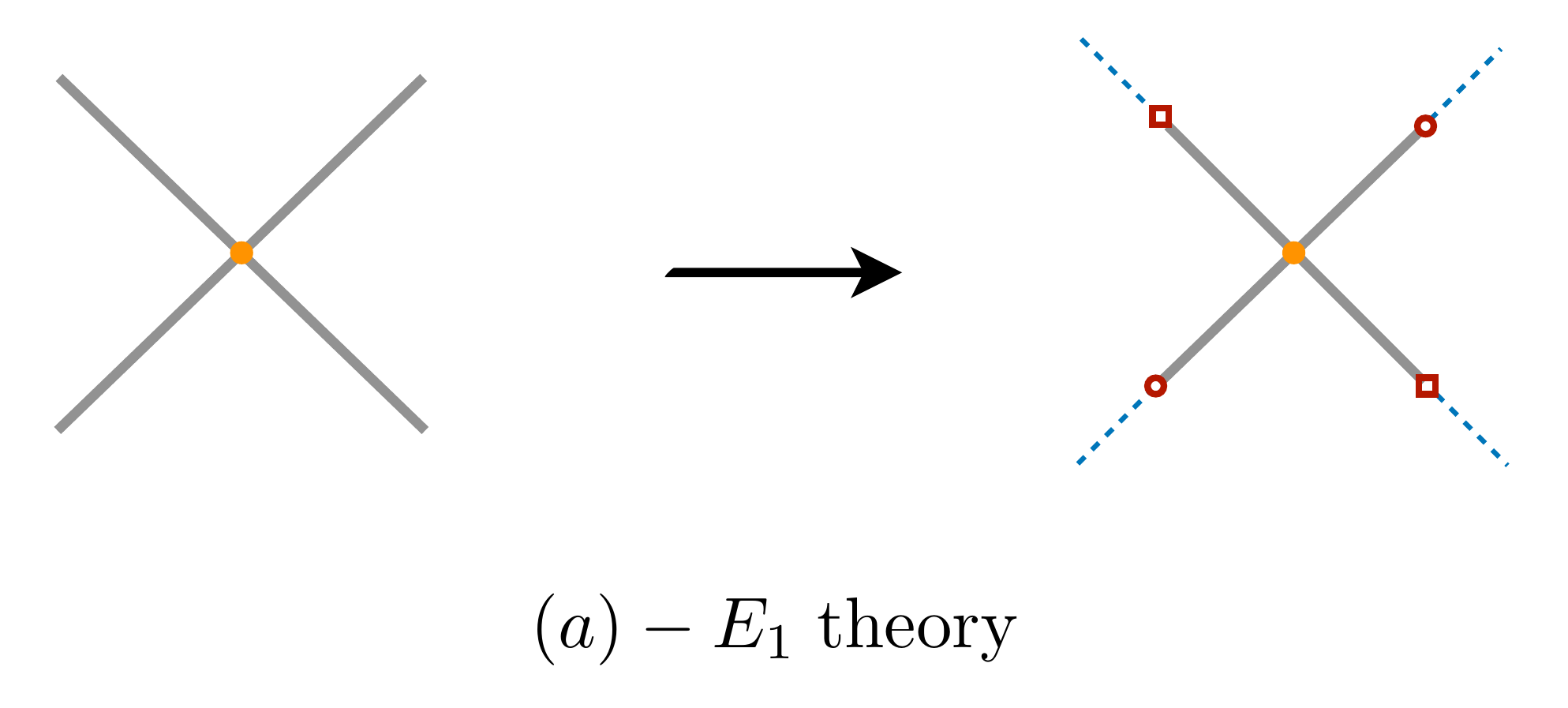}
	\includegraphics[scale=0.27]{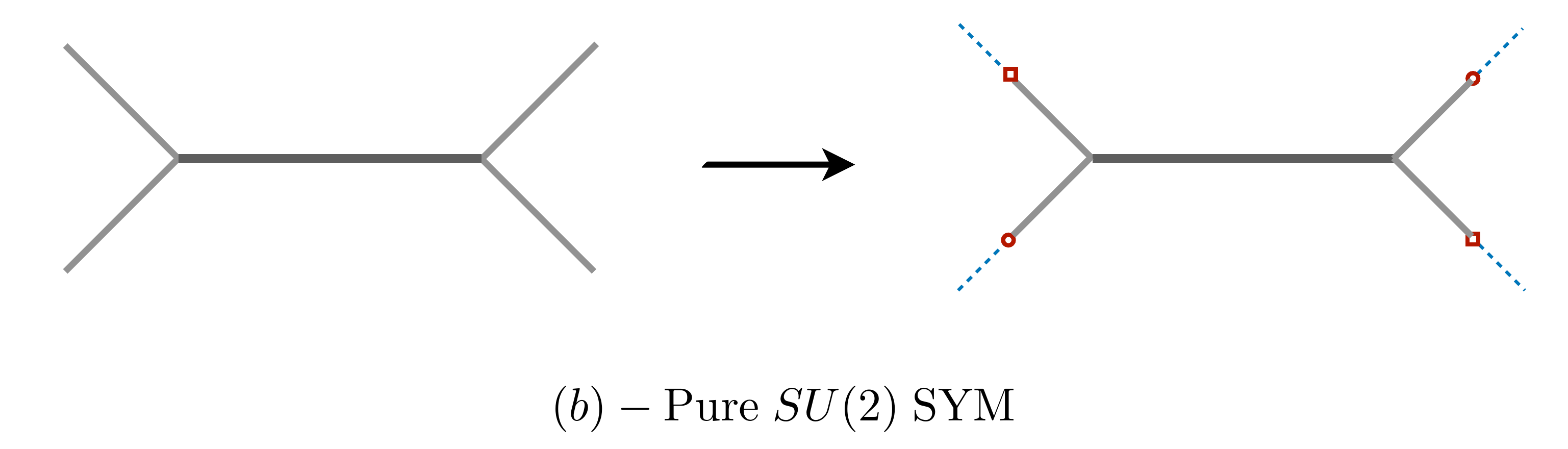}
	\caption{The description of the $E_1$ theory and of pure SU(2) SYM with 7-branes in. Red circles/squares are [1,1] and [1,-1] 7-branes, respectively, which are localized on the $(x,y)$ plane. Dashed blu lines represent 7-brane branch cuts.}
	\label{pq_7}
\end{figure}

As shown in \cite{DeWolfe:1999hj}, 5-branes of type (p,q) can end on [p,q] 7-branes. All semi-infinite branes in the web can be then made finite by letting them end on a 7-brane of the same type. To preserve supersymmetry, the latter should be transverse to the $(x,y)$ plane. Figure \ref{pq_7} provides the dictionary between the descriptions of the $E_1$ theory and pure $SU(2)$ SYM  with/without 7-branes.

\begin{figure}[h!]
	\centering
	\includegraphics[scale=0.32]{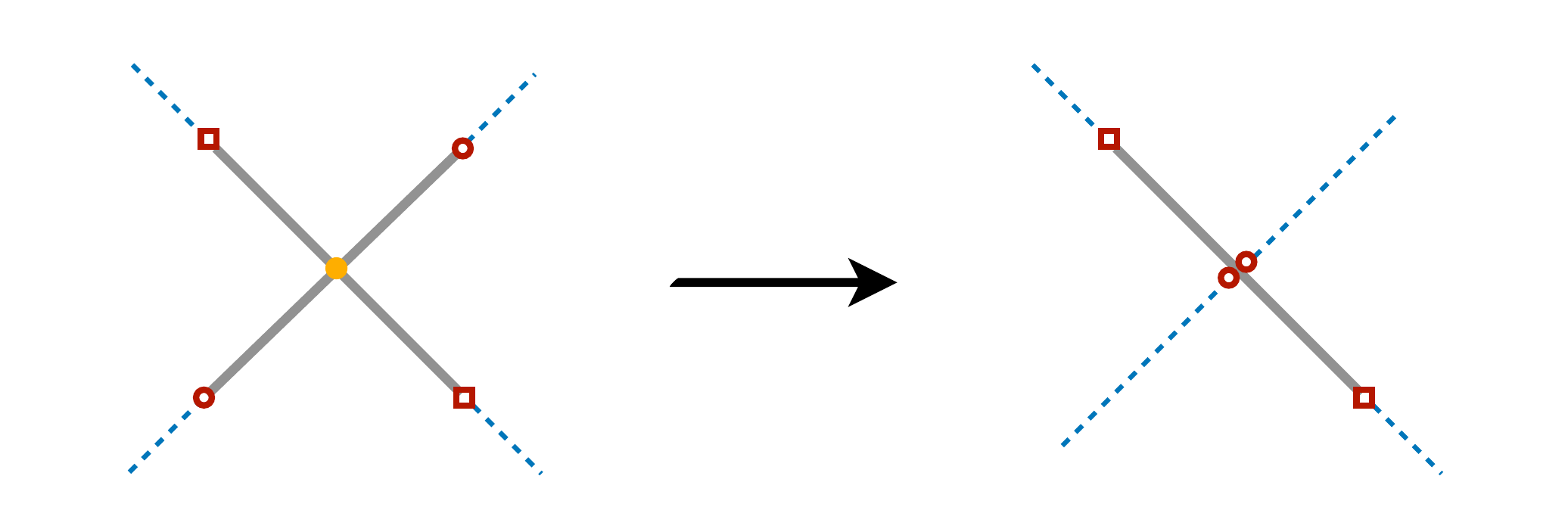}
	\caption{Instanton symmetry enhancement. The [1,1] 7-branes can be moved along the 5-brane prong without changing in any way the 5d theory. Hence the two pictures above describe the same physics. The one of the right makes it manifest the enhancement of the instantonic symmetry $U(1)_I \rightarrow SU(2)_I$ at the $E_1$ fixed point.}
	\label{pq_8}
\end{figure}

The 7-brane gauge group acts as a global symmetry on the 5-brane theory, and in this set-up the flavor symmetry enhancement at the UV fixed point becomes manifest. Indeed, 7-branes can be moved at no cost, i.e. without affecting the 5d theory, along the 5-branes ending on them  \cite{DeWolfe:1999hj}. Hence, at infinite coupling, figure \ref{pq_7}($a$), two $[1,1]$ 7-branes can be brought to the same point on the $(x,y)$ plane (equivalently two $[1,-1]$ 7-branes, due to the obvious symmetry of the brane web). When the two are on top of each other the $U(1)_I$ global symmetry gets enhanced to $SU(2)_I$, as illustrated in figure \ref{pq_8}. There cannot  be any further symmetry enhancement since two $[1,1]$ branes cannot be put on top of a $[1,-1]$ 7-brane \cite{DeWolfe:1999hj,DeWolfe:1998zf}. Hence only two out of four, of either $[1,1]$ or $[1,-1]$ types, can be superimposed. Note that this operation cannot be done at finite $1/g^2$, as it is clear from figure \ref{pq_7}($b$), since the minimal distance the two $[1,1]$ 7-branes can reach is in fact $h=1/g^2$. This shows that only at infinite coupling symmetry enhancement occurs, in agreement with field theory expectations. 

\subsection{Moduli space and supersymmetric deformations}
\label{branches}
 
Let us now focus on the moduli space of the $E_1$ theory and on its supersymmetric deformations. We first summarize known field theory results and then show how they can be described in (p,q)-brane web brane language.

The $E_1$ theory admits both a Coulomb branch and a Higgs branch. The former is present also in ${\cal N}=1$ $SU(2)$ pure SYM while the latter is peculiar of the $E_1$ theory only. The Higgs branch is a one-dimensional hyper-K\"ahler sigma model $\mathbb{C}^2/\mathbb{Z}_2$ parametrized, locally, by a complex hypermultiplet ${\cal H}$ with scalar components $H_i$,  where $i$ is an index in the fundamental of $SU(2)_R$.  At a generic point on the Higgs branch the global symmetry is broken to  $D[SU(2)_I \times SU(2)_R] = SU(2)_{R'}$, which is still an R-symmetry. This gives rise to three Goldstone modes which, together with the scaling modulus, describe the one-dimensional quaternionic sigma model.

This parametrization hides the action of the instantonic symmetry on the Higgs branch and of the aforementioned symmetry breaking pattern. To make this action manifest, the complex hypermultiplet can more conveniently be described as two real half hypers ${\cal H}^a$ with $a$ an index in the fundamental representation of $SU(2)_I$ and ${\cal H}^a = \overline{\cal H}_a$.

Using real half hyper parametrization, one can easily see the lifting of the Higgs branch at finite gauge coupling. This is done introducing background vector fields of the instantonic  $SU(2)_I$ symmetry. In five dimensions the vector multiplet has the following structure  
\begin{equation}
\label{vecmb}
{\cal V}_{(ab)} = \left(\phi_{(ab)} , A^\mu_{(ab)} , \lambda^i_{(ab)} ,  Y^{(ij)}_{(ab)} \right)~,
\end{equation}
where $Y^{(ij)}_{(ab)}$ is a $SU(2)_R$ triplet of scalar fields and $\phi_{(ab)}$ a real scalar. All fields in \eqref{vecmb} transform in the adjoint of $SU(2)_I$. This (background)  vector multiplet acts as a source for the linear $SU(2)_I$ current multiplet ${\cal J}^{(ab)}$ of the $E_1$ theory.  Its components are \cite{Bhattacharya:2008zy,Tachikawa:2015mha,Cordova:2016xhm}
\begin{equation}
\label{lcurr1}
{\cal J}^{(ab)} = \left(\mu_{(ij)}^{(ab)} , \psi^{(ab)}_m , J^{(ab)}_\mu ,  M^{(ab)}\right)
\end{equation}
where besides the current there are a fermionic operator $\psi^{(ab)}_m$ and two scalar operators, $\mu_{(ij)}^{(ab)}$ and $M^{(ab)}$.  The corresponding scaling dimensions are
\begin{equation}
\label{lcurr2}
\Delta(\mu_{(ij)}^{(ab)})=3 ~,~ \Delta(\psi^{(ab)}_m)=7/2 ~,~ \Delta(J^{(ab)}_\mu)=4 ~,~  \Delta(M^{(ab)})=4~.
\end{equation}
The $SU(2)_I$ current multiplet can be described in terms of the real half hypermultiplets ${\cal H}^a$. 
The hypermultiplet scalars $H^a_i$ transform in the bi-fundamental of $SU(2)_I \times SU(2)_R$ global symmetry. Since the $SU(2)_I$ instantonic symmetry acts on the Higgs branch, and hence on  $H^a_i$, the bottom component $\mu_{(ij)}^{(ab)}$ of the $SU(2)_I$ current multiplet can reliably be tracked on the Higgs branch:\footnote{We thank Thomas Dumitrescu for clarifying this to us.} it is simply the tri-holomorphic moment map of the $SU(2)_I$ symmetry and reads 
\begin{equation}
\label{current1}
\mu_{(ij)}^{(ab)} \sim H^{(a}_{(i} H^{b)}_{j)}~,
\end{equation}
which in turns implies, for instance, that 
\begin{equation}
\label{current2}
M^{(ab)} \sim  \Omega^{\alpha \beta} Q_\alpha^i Q_\beta^j \mu^{(ab)}_{(ij)}~,
\end{equation}
where $ \Omega^{\alpha \beta}$ is the $Spin(5)$ symplectic invariant tensor and $Q_\alpha^i $ are the supercharges.

${\cal N}=1$ supersymmetry couples the background gauge field with the $SU(2)_I$ current multiplet ${\cal J}^{(ab)}$ and by \eqref{current1} and \eqref{current2} to  the hyperscalars $H_i^a$  as
\begin{equation}
\label{gauge7}
Y^{(ij)}_{(ab)} \cdot  H^{(a}_{(i} H^{b)}_{j)}~~~\mbox{and}~~~ \phi_{(ab)}\cdot \Omega^{\alpha \beta} Q_\alpha^i Q_\beta^j H^{(a}_{(i} H^{b)}_{j)}~. 
\end{equation}
A VEV for the scalar operator $\phi_{(ab)}$ corresponds to a massive deformation proportional to the inverse gauge coupling squared, $1/g^2$  \cite{Seiberg:1996bd}. This brings the $E_1$ theory down to $SU(2)$ SYM and breaks the global $SU(2)_I$ symmetry to $U(1)_I$ while leaving the $SU(2)_R$ symmetry untouched, as expected for ${\cal N}=1$ $SU(2)$ SYM. 

In fact, via \eqref{gauge7}, $\langle \phi_{(ab)} \rangle $ provides a mass also to the hypermultiplets ${\cal H}^a$ 
\begin{equation}
\label{susyc}
\phi_{(ab)}\, M^{(ab)} ~ \longrightarrow ~ \Delta {\cal L} = \langle \phi_{(12)} \rangle M^{(12)} = \frac{1}{g^2}  M^{(12)} = \frac{1}{g^2} \Omega^{\alpha \beta} Q_\alpha^i Q_\beta^j \, H^{(1}_{(i} H^{2)}_{j)}~,
\end{equation}
where we have chosen a specific Cartan direction, for definiteness. We hence see that the Higgs branch is lifted at weak coupling, as anticipated. This is a supersymmetric mass deformation since $M^{(ab)}$ is a highest component operator of the $SU(2)_I$ current multiplet and so the full hypermultiplet becomes massive, i.e. both $H^a_i$ and its fermionic superpartner $\psi^a_H$, with a mass $m_{\mbox{\tiny SUSY}}=1/g^2$.

The $E_1$ theory and $SU(2)$ SYM admit also a Coulomb branch. This is parametrized in terms of VEV of a  scalar operator $\sigma$ (which at weak coupling corresponds to the scalar component in the Cartan of the $SU(2)$ vector multiplet). This scalar operator does not carry any global index, hence along the Coulomb branch the full global symmetry $SU(2)_I \times SU(2)_R$ (or $U(1)_I \times SU(2)_R$ at finite coupling) is preserved. Further, the hypermultiplet parametrizing the Higgs branch gets a mass proportional to $\langle \sigma \rangle$, showing that the $E_1$ theory does not admit a mixed branch. One reason for this to be expected is that the Higgs branch is singular at its origin and it is known that the Coulomb branch  of the $E_1$ theory is smooth for any finite value of $\langle \sigma \rangle$ \cite{Seiberg:1996bd,Morrison:1996xf}. Another way to see that the Higgs branch is lifted on the Coulomb branch comes from the BPS bound for instantonic particles. At weak coupling, the breaking of the $SU(2)$ gauge symmetry on the Coulomb branch generates a CS level $k$ for the Cartan surviving the breaking. The effect of this CS level consists in giving a net additional electric charge $n_e=k I$ to particles charged under the instantonic symmetry, with $I$ the $U(1)_I$ charge \cite{Seiberg:1996bd}. If the particle is BPS, its mass then reads
\begin{equation}
\label{BPS_eq}
M=|n_e\langle \sigma \rangle +h I |~.
\end{equation}
Since this bound remains true also in the strong coupling limit, particles charged under the instantonic symmetry, as the hypermultiplet ${\cal H}^a$, obtain a mass proportional to $\langle \sigma \rangle$ on the Coulomb branch \cite{Aharony:1997bh}.

\subsubsection*{Brane web description}
\label{branches_pq}

Let us now see how what we discussed so far is realized using 5-brane web brane language.  As originally discussed in \cite{Aharony:1997bh,DeWolfe:1999hj}, the Higgs and Coulomb branches of the theory are described by local deformations of the brane web (i.e. deformations that do not change the position of the external 5-branes in the $(x,y)$ plane). Global deformations of the brane web, i.e. changes in the position of external 5-branes in the $(x,y)$ plane, correspond instead to actual deformations of the field theory -- an example being the change in the gauge coupling $g$, see figures \ref{pq_1} and \ref{pq_7}.

\begin{figure}[h!]
	\centering
	\includegraphics[scale=0.32]{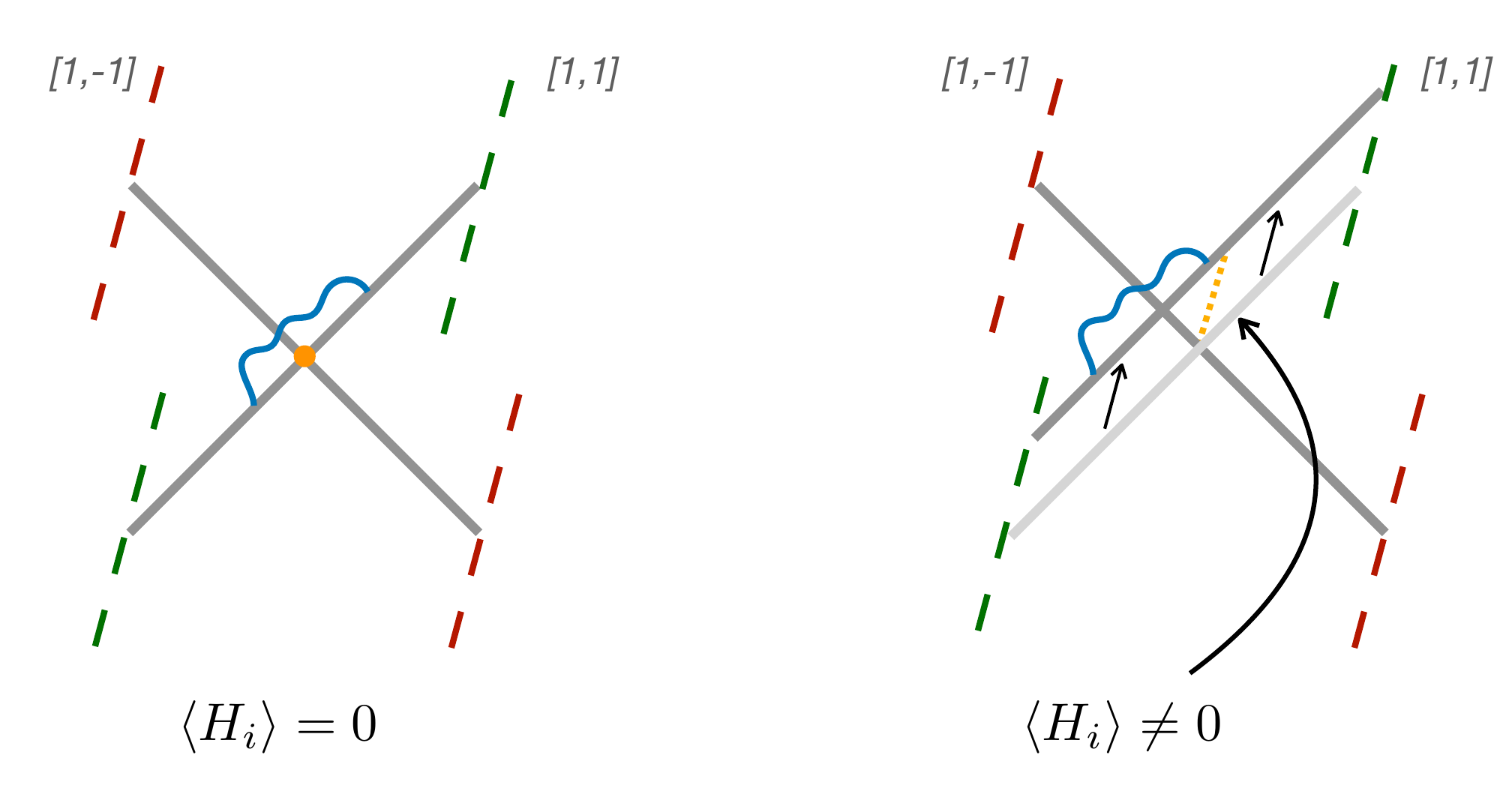}
	\caption{The $E_1$ theory at the origin (left) and at an arbitrary point (right) of the Higgs branch. The blue wiggles represent the hypermultiplet ${\cal H}$. The $(1,1)$ 5-branes slides along the $[1,1]$ 7-branes, giving rise to a free hypermultiplet (a finite 5-brane stretching between two 7-branes). The VEV of $H_i $ is proportional to the distance between the $(1,1)$ and $(1,-1)$ 5-branes in the transverse space. 
	}
	\label{pq2_7b}
\end{figure}

Let us start by considering the Higgs branch, which is a property of the $E_1$ theory, only. This is described in figure \ref{pq2_7b} and it is obtained by displacing the $(1,1)$ 5-brane in the space transverse to the $(x,y)$ plane.   
The complex hypermultiplet ${\cal H}$ parametrizing the Higgs branch can be identified following \cite{Hanany:1996ie,Cherkis:2008ip}. It corresponds to the lowest energy excitations of the $(1,1)$ strings on the $(1,1)$ 5-branes across the $(1,-1)$ 5-branes, which is indeed a free hypermultiplet when the $(1,1)$ and $(1,-1)$ 5-branes do not intersect \cite{Benini:2009gi}.

This process describes a symmetry breaking pattern $SU(2)_R \rightarrow U(1)_R$: the brane separation occurs in the transverse directions of the 5-brane system, which is $\mathbb{R}^3$, and hence only $SO(2)$ rotations, corresponding to a $U(1)_R$ symmetry, survive in the transverse space. It is worth noticing that one cannot display the full Higgs branch symmetry breaking pattern $SU(2)_I \times SU(2)_R \rightarrow SU(2)_{R'}$. This is because the two symmetries, the R-symmetry and the instantonic symmetry, have different realizations in the brane web. The former is realized geometrically, in terms of rotations in the three-dimensional transverse space the 5-brane system shares. The latter is instead realized in terms of string degrees of freedom, namely the (1,1) strings living on the [1,1] 7-branes. These two symmetries have different origin and cannot mix, so the web diagram is not manifestly invariant under the unbroken group.

\begin{figure}[h!]
	\centering
	\includegraphics[scale=0.31]{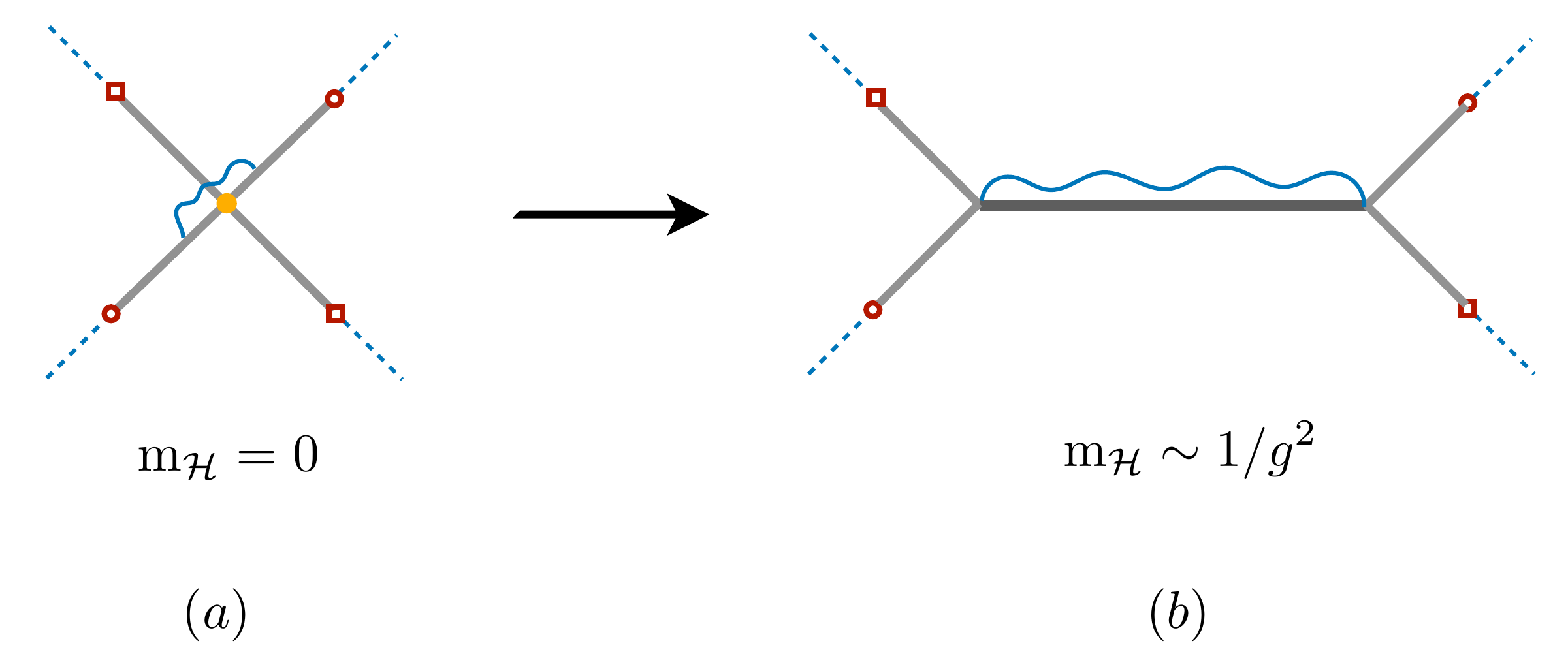}
	\caption{Higgs branch lifting and (a naive description of) hypermultiplet mass generation at finite gauge coupling.}
	\label{pq_3}
\end{figure}

Let us now consider the supersymmetric mass deformation \eqref{susyc} which makes the $E_1$ theory flow to pure $SU(2)$ SYM. This is described by the global deformation depicted in figure \ref{pq_3}. We see from the figure that the Higgs branch is lifted, as expected. Indeed,  as soon as $1/g^2 \not =0$  the $(1,1)$ and $(1,-1)$ 5-branes cannot anymore be moved apart since the quadruple brane junction splits into two triple ones. At the same time, the $(1,1)$ strings describing the hypermultiplet ${\cal H}$ get stretched and get a mass proportional to $1/g^2$, in agreement with field theory expectations. 

An important comment is worth at this point. Although from figure \ref{pq_3}($b$) we do see that ${\cal H}$ gets a mass at finite gauge coupling, this description is inaccurate. The hypermultiplet parametrizing the Higgs branch is a BPS state but  a $(1,1)$ string connecting two separated $(1,1)$ branes as in figure \ref{pq_3}($b$) is not. A similar problem arises when adding flavors on a brane web\cite{Aharony:1997bh,DeWolfe:1999hj}. As in that case, a simpler and more direct description can be obtained performing a Hanany-Witten (HW) transition \cite{Hanany:1996ie} on the original brane web, as described in figure \ref{pq3_par2} (see the appendix of \cite{Bergman:2020myx} for a thorough discussion of HW transitions in this context).  After the transition, a $(1,1)$ string stretching between the $(1,1)$ 5-brane and the $[1,1]$ 7-brane can be constructed and correctly represents the hypermultiplet degrees of freedom. 
\begin{figure}[h!]
	\centering
	\includegraphics[scale=0.35]{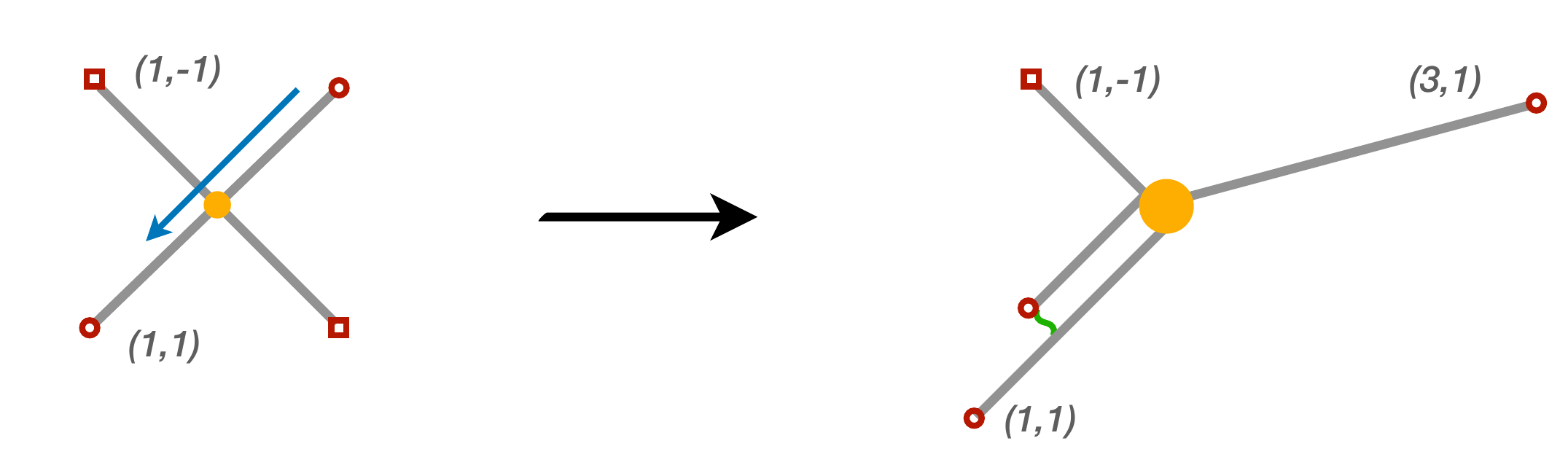}
	\caption{Hanany-Witten transition. In the resulting 5-brane web the string representing the hypermultiplet is now a $(1,1)$ string stretched between the $(1,1)$ 7-branes and 5-branes (green wiggle). The two (1,1) 5-branes on the bottom left are displaced just to visualize the hypermultiplet degrees of freedom.}
	\label{pq3_par2}
\end{figure}

\begin{figure}[h!]
	\centering
	\includegraphics[scale=0.35]{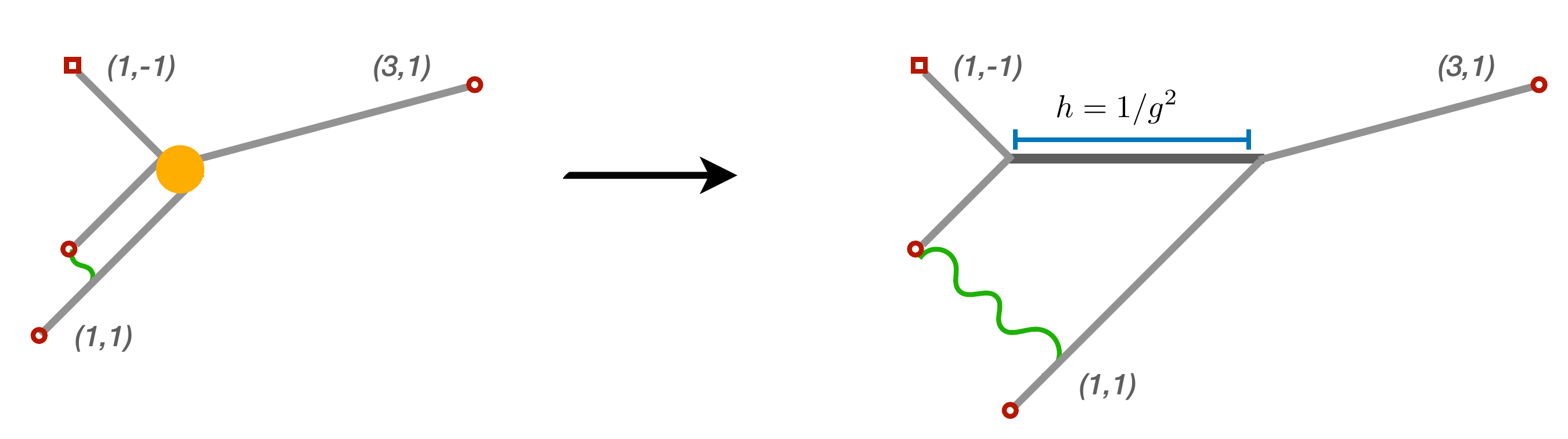}
	\caption{Supersymmetric mass deformation. The $(1,1)$ string representing the hypermultiplet ${\cal H}^a$ gets a mass proportional to the inverse coupling squared. }
	\label{pq3_par3}
\end{figure}

 As illustrated in figure  \ref{pq3_par3}, at finite gauge coupling this string acquires a mass which, in units of the fundamental string tension $T_s$, is
\begin{equation} 
\label{posmh}
m_{\cal H} = 1/g^2~,
\end{equation} 
which is the same of an instanton with unit charge. As it is clear from the figure, the $(1,1)$ string length is suppressed by a $\sqrt{2}$ factor with respect to that of the D1-brane on the D5-brane which is an instanton of the $SU(2)$ theory, but its tension is $\sqrt{2}$ larger and the two factors compensate. This shows that this state is BPS and at threshold with an instanton with $I=1$, see eq.~\eqref{BPS_eq}.\footnote{This agrees with the fact that all BPS saturated states in the ${\cal N}=1$ $SU(2)$ theory can be viewed as bound states of two basic states, an instanton and, when the Coulomb branch opens-up, a massive W-boson (see \cite{Aharony:1997bh} for a detailed discussion).}

\begin{figure}[h!]
	\centering
	\includegraphics[scale=0.35]{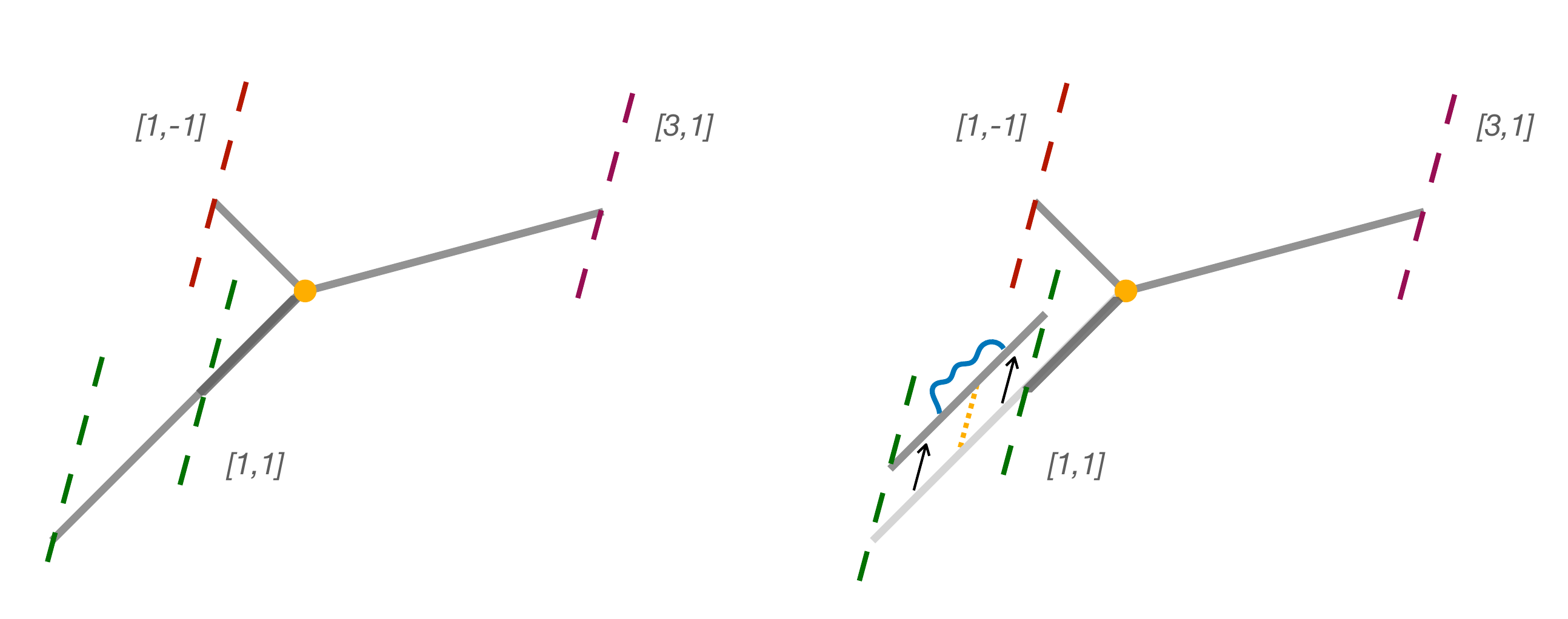}
	\caption{Higgs branch description in the HW dual frame. Displacing a finite  $(1,1)$ 5-brane in the transverse space provides a VEV to the hyperscalars $H^a_i$ \cite{Cherkis:2008ip}. The blue wiggles on the 5-brane describe the complex hypermultiplet parametrizing the Higgs branch.}
	\label{pq3_par4}
\end{figure}

In this dual frame the hypermultiplet degrees of freedom are naturally described in terms of the real half hypermultiplet ${\cal H}^a$, which are nothing but the $(1,1)$ strings connecting 5 and 7-branes in figure \ref{pq3_par2}. The Higgs branch is obtained by displacing a $(1,1)$ 5-brane segment in the transverse space, as shown in figure \ref{pq3_par4}. The free hyper parametrizing the Higgs branch are $(1,1)$ strings living on the finite $(1,1)$ 5-brane (blue wiggles in the figure), which is in fact a complex hyper, as in figure \ref{pq2_7b}.  

\begin{figure}[h!]
	\centering
	\includegraphics[scale=0.3]{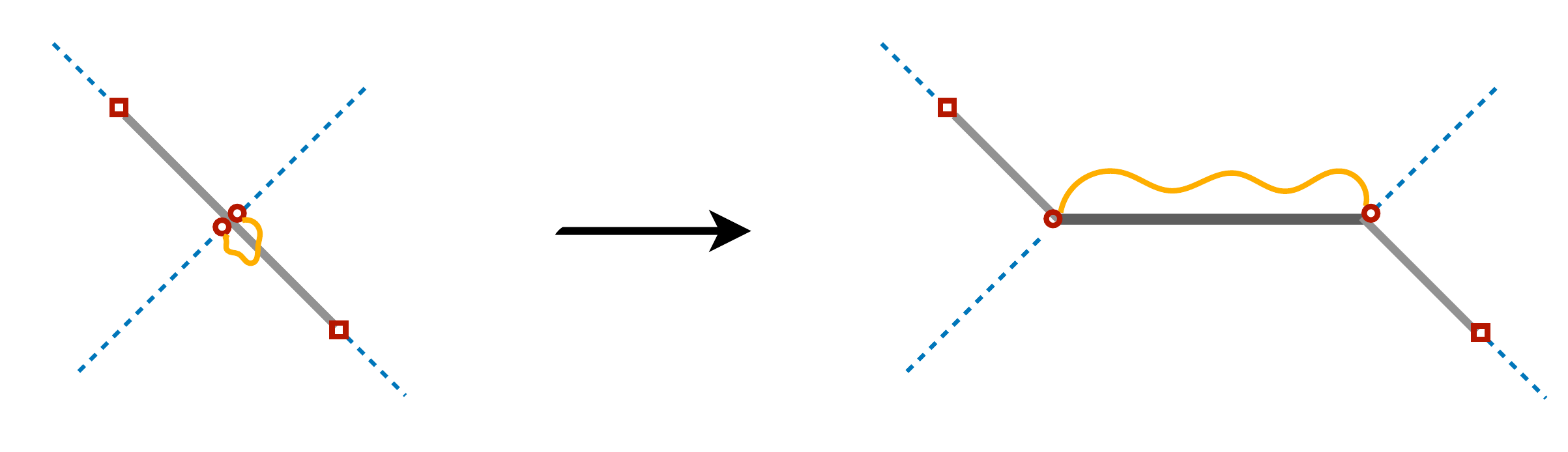}
	\caption{Geometric description of the symmetric breaking $SU(2)_I \rightarrow U(1)_I$ induced by the mass deformation \eqref{susyc}. At $1/g^2 \not =0$  the two $[1,1]$ 7-branes cannot anymore be put on top of eachother, the $(1,1)$ strings living on them (yellow wiggles) get stretched  and the instantonic symmetry is broken to $U(1)_I$. The R-symmetry remains untouched.}
	\label{pq_11}
\end{figure}

Let us now see how the brane web captures the correct symmetry breaking pattern induced by the mass deformation. From figure \ref{pq_3} one can also see the correct symmetry breaking pattern $SU(2)_I \times SU(2)_R \rightarrow U(1)_I \times SU(2)_R$. 
From the brane web perspective, the $SU(2)_I$ group is realized as the gauge group of the 7-branes. The low energy excitations of the $(1,1)$ strings living on the $[1,1]$ 7-branes are a vector multiplet in eight dimensions. Once reduced to 5 dimensions, in terms of $\mathcal{N}=1$ representations this corresponds to a (free) hypermultiplet and a vector multiplet (at zero coupling). The latter represents the background vector multiplet of the flavor symmetry of the five-dimensional $E_1$ theory, eq.~\eqref{vecmb}. At finite gauge coupling the $[1,1]$ 7-branes get displaced due to the finite length $(2,0)$ 5-brane and the $(1,1)$ strings connecting the 7-branes get stretched and acquire a minimal length of order $1/g^2$, see figure \ref{pq_11}. This is Higgsing for the $SU(2)_I$ theory living on the $[1,1]$ 7-branes, which is broken to $U(1)_I$, and corresponds to give a VEV $\sim 1/g^2$ to the lowest component of the (background) vector multiplet, $\phi_{(ab)}$. The $SU(2)_R$ symmetry is instead preserved since the deformation does not involve the transverse space.\footnote{An identical argument can be done in the dual setup of figure \ref{pq3_par3}.} 

\begin{figure}[h!]
	\centering
	\includegraphics[scale=0.25]{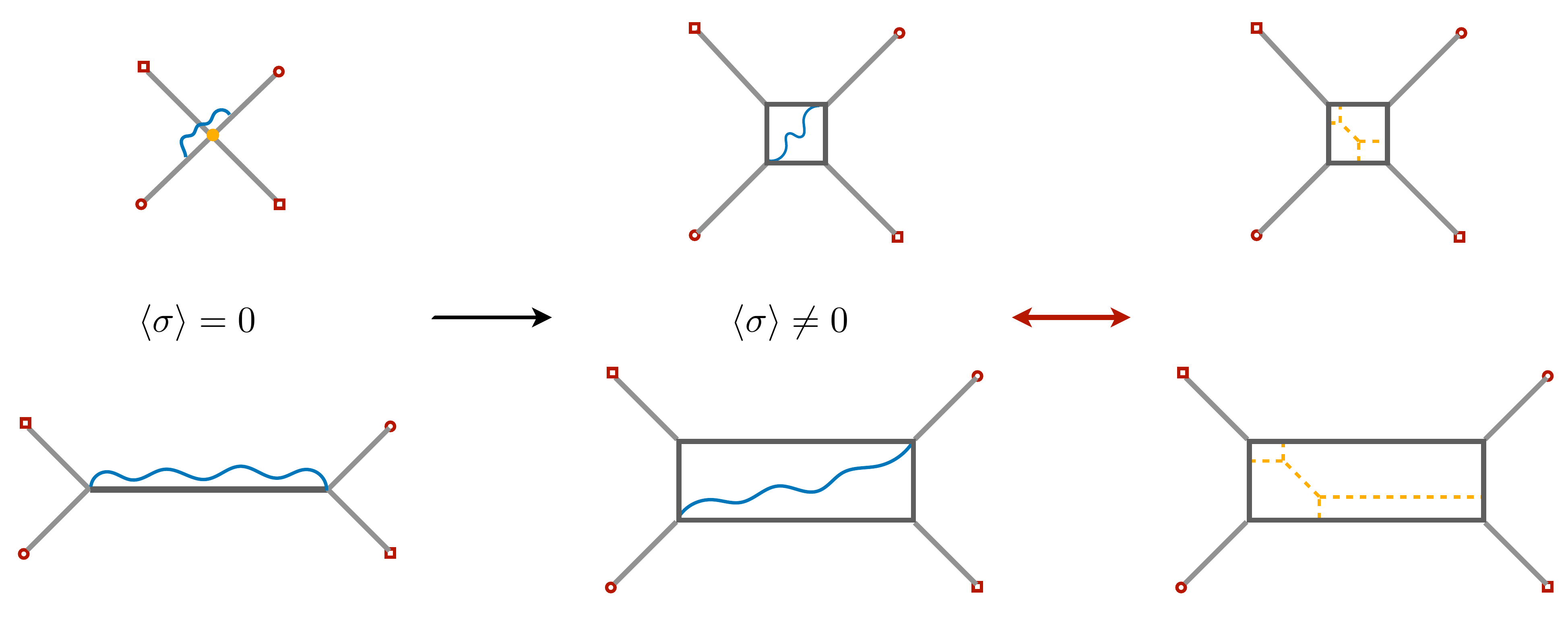}
	\caption{Coulomb branch description for the $E_1$ theory (up) and at weak coupling (down). The $E_1$ theory Higgs branch is lifted since the (1,1) 5-brane cannot anymore be displaced in the transverse space. The hypermultiplet (blue wiggles in the figure) becomes massive. The corresponding BPS state has $n_e=2$ and (at finite gauge coupling - bottom figures) instantonic charge $I=1$. It is at threshold with a bound state of an instanton and a W-boson, whose (p,q)-brane web description is depicted in the most right figures.}
	\label{pq_3b}
\end{figure}

Let us finally discuss the Coulomb branch, which exists both for the $E_1$ theory and for ${\cal N}=1$ $SU(2)$ SYM. We report in figure \ref{pq_3b} its description both at weak coupling and at infinite coupling. As for the latter, one easily see that the Higgs branch is lifted, since the $(1,1)$ 5-brane cannot anymore be moved apart along the transverse space, in agreement with field theory analysis.  
The $(1,1)$ strings describing the hypermultiplet ${\cal H}$ becomes massive and saturates the BPS mass formula \eqref{BPS_eq} with $n_e=2$ and $I=1$. It is then at threshold with a bound state of a W-boson ($n_e=1 , I=0$) and an instanton with unit instantonic charge ($n_e=1, I=1$ -- recall that on the Coulomb branch an instanton acquires an effective $U(1)$ charge \cite{Aharony:1997bh}). 

\begin{figure}[h!]
	\centering
	\includegraphics[scale=0.28]{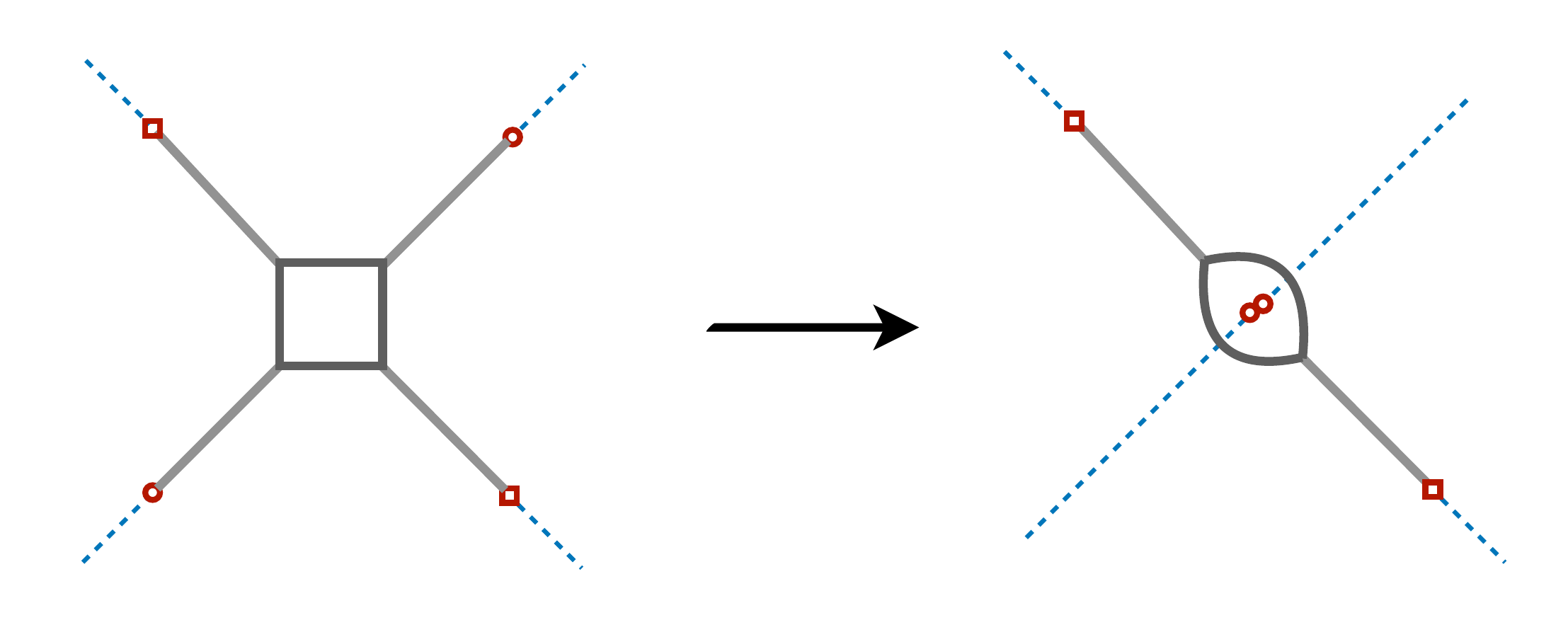}
	\caption{On the Coulomb branch of the $E_1$ theory the $[1,1]$ 7-branes can still be moved on top of eachother and the full $SU(2)_I$ global symmetry is preserved. The presence of 7-branes creates a non-trivial metric so the 5-branes follow curved geodesics. The curved 5-branes in the right figure is just a pictorial way to mimic this effect (which does not change the nature of the 5-branes, namely their charges).}
	\label{pq_12}
\end{figure}

Finally, in agreement with field theory expectations, one can see that on the Coulomb branch the full global symmetry is preserved, $SU(2)_ I \times SU(2)_R$ and $U(1)_I \times SU(2)_R$ at infinite and finite couplings, respectively. The $SU(2)_R$ symmetry is preserved since the deformation does not involve  transverse directions, as opposed to the Higgs branch deformation described in figure \ref{pq2_7b}. Also the instantonic symmetry is preserved. In particular, as shown in figure \ref{pq_12}, in the $E_1$ theory the $[1,1]$ 7-branes can still be freely moved on the $(1,1)$ 5-branes prong, preserving the full $SU(2)_I$ symmetry. 

\section{Supersymmetry breaking mass deformation}
\label{susybr}
In this section we discuss the supersymmetry breaking deformation proposed in \cite{BenettiGenolini:2019zth}, provide its geometric realization via brane webs and discuss its effects on the low energy dynamics. 

Starting from the $E_1$ fixed point one can consider a more general deformation than \eqref{susyc}, in which also the lowest component operator of the $SU(2)_I$ current multiplet $\mu^{(ab)}_{(ij)}$ is activated. Being a lowest component operator, this operation breaks supersymmetry. Properly aligning the Cartan generators of $SU(2)_I$ and $SU(2)_R$ and the operators $M^{(ab)}$ and $\mu^{(ab)}_{(ij)}$ accordingly,  this more general deformation can be written as
\begin{equation}
\label{fulldef}
\Delta {\cal L} = h M^{(12)} + \widetilde m \mu^{(12)}_{(12)}~,
\end{equation}
where $\Delta(h)=1$ and $\Delta(\widetilde m)=2$. For large $h$ or, more precisely, in the regime $|\widetilde m| <<h^2$ one can look at this deformation as a small supersymmetry breaking deformation of ${\cal N}=1$ $SU(2)$ SYM. In this regime one can easily show that under the more general deformation \eqref{fulldef} both the gaugini and the scalar field $\sigma$ belonging to the $SU(2)$ gauge multiplet get a mass proportional to $\widetilde m$ and the low energy dynamics reduces to pure $SU(2)$ YM.\footnote{For later purpose let us notice that this also means that the Coulomb branch is lifted.} One could wonder what happens instead at strong coupling or, say, in the regime $|\widetilde m| >> h^2$. 

As discussed in \cite{BenettiGenolini:2019zth}, that something different should happen could be inferred by coupling the theory to background gauge fields for the $U(1)_I \times U(1)_R$ global symmetry and compute the corresponding CS levels 
\begin{equation}
{\cal S}_{CS} = \frac{k_a}{24\pi^2}\int A_a \wedge dA_a \wedge dA_a~~~,~~~a=I,R~.
\end{equation}
Explicit computations give a scenario as in figure \ref{ps_0}, with $k_I=-2,k_R=-3/2$. This means that the CS levels jump when crossing the axes. While the jump across the $h$-axis can be understood in terms of gaugini becoming massless there (supersymmetry is restored at $\widetilde m=0$), no such explanation holds in crossing the $\widetilde m$-axis, where the jump in the CS levels occurs also for the topological $U(1)_I$ symmetry, under which perturbative states are neutral. So, on the $\widetilde m$-axis, or in its vicinity, say in the regime $|\widetilde m| >> h^2$, where perturbative intuition does not help, something should indeed happen. 

In order to see what that can be let us look at the supersymmetry breaking deformation \eqref{fulldef} more closely and start considering to deform the $E_1$ fixed point just with
\begin{equation}
\label{halfdef}
\Delta {\cal L} = \widetilde m \mu^{(12)}_{(12)}~,
\end{equation}
namely at infinite coupling, $h=0$. One can easily show that this deformation induces an instability on the Higgs branch, which in the $E_1$ theory is instead a flat direction. More precisely, under \eqref{halfdef} a mass term is generated for the hyperscalars $H^a_i$ parametrizing the Higgs branch but this mass term always has a positive and a negative eigenvalues. To see this, let us rewrite \eqref{halfdef}  using \eqref{current1}  as
\begin{equation}
\label{halfdef1}
\Delta {\cal L} = \widetilde m \left(H^1_1 H^2_2 + H^2_1 H^1_2\right)~.
\end{equation}
Recalling that $H^{ia} = \overline{H}_{ia}$, with little algebra the above equation can be recasted as
\begin{equation}
\label{halfdef2}
\Delta {\cal L} = \widetilde m \left(|H^1_1|^2 - |H^2_1|^2\right)~,
\end{equation}
which shows that regardless the sign of $\widetilde m$ there always exists a tachyonic mode and hence an instability.\footnote{We thank Thomas Dumitrescu for pointing this out to us.} Considering the more general deformation \eqref{fulldef} we then see that there are two competitive contributions to scalar masses, one weighted by $\widetilde m$ and one by $h$ (the latter, being a supersymmetric mass deformation, always gives positive mass squared contribution for the hyperscalars $H^a_i$) 
\begin{equation}
\label{halfdef3}
\Delta {\cal L} = (h^2 + \widetilde m) |H^1_1|^2  + (h^2 - \widetilde m) |H^2_1|^2 ~.
\end{equation}
This implies that, unlike the regime $|\widetilde m| << h^2$, the regime $|\widetilde m| >> h^2$ describes a region of instability. 
Before discussing what the fate of this instability could be, let us see how this discussion translates in brane-web language.

\vskip 10pt

\begin{figure}[h!]
	\centering
	\includegraphics[scale=0.43]{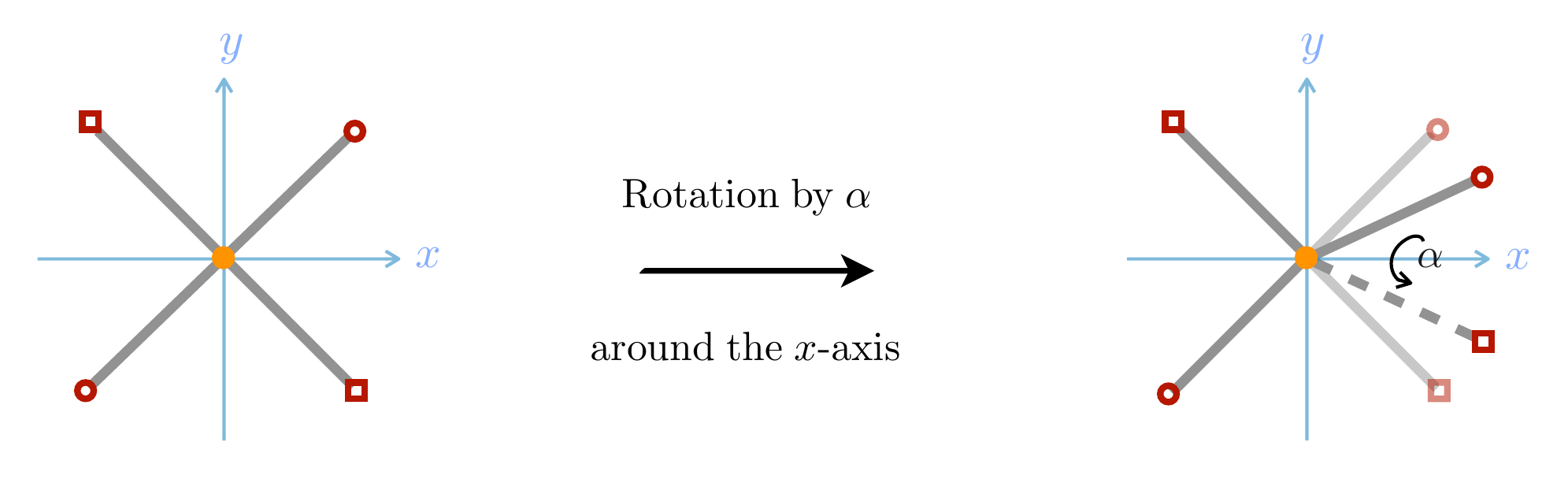}
	\caption{The supersymmetry breaking deformation: a rigid rotation around the $x$-axis of the right (1,1) and (1,-1) 5-branes.}
	\label{pq_4}
\end{figure}

The supersymmetry breaking deformation \eqref{halfdef} has (at least) the following properties: 
\begin{enumerate}
\item Break supersymmetry.
\item Break $SU(2)_I \times SU(2)_R$ to $U(1)_I \times U(1)_R$.
\item Lift the Higgs branch of the $E_1$ theory.
\item Lift the Coulomb branch. 
\end{enumerate}
Starting from the $E_1$ theory, figure \ref{pq_7}($a$), let us consider a rigid rotation of the two (semi-infinite) right 5-branes along the $x$-axis, as shown in figure \ref{pq_4}.

This deformation satisfies all above requirements. First, the rotation along the $x$-axis changes the angles between the 5-branes of the $E_1$ fixed point and therefore the supersymmetric condition for brane junctions is not anymore satisfied. Supersymmetry is then broken. The R-symmetry is broken to $U(1)_R$ since only $SO(2)$ rotations are still allowed in the transverse space. The rotation by an angle $\alpha$ also misaligns the [1,1] 7-branes on which the (1,1) 5-branes end which get rotated one another by the very same angle. Therefore, the 7-branes are not anymore parallel and the $SU(2)_I$ instantonic symmetry is broken to $U(1)_I$.\footnote{This also signals the instability. The tilt, besides higgsing the $SU(2)_I$ gauge theory on the 7-branes, makes open strings connecting the two 7-branes becoming tachyonic, one mode getting a positive mass squared and the other one getting a negative mass squared  \cite{Sen:1999mg,Aldazabal:2000dg}. This can be seen as the D-term of the $SU(2)_I$ vector multiplet on the 7-branes getting a VEV along a Cartan direction.} The Higgs branch is lifted since the possibility to displace any of the 5-branes along transverse directions at no cost, as in figure \ref{pq2_7b}, is now geometrically obstructed. Finally, also the Coulomb branch is lifted. The four 5-branes describing the string junction cannot undergo a process as the one described in figure \ref{pq_3b} because the finite length D5 and NS5 branes would now be misaligned and it costs energy to open up the string junction. So the modulus $\sigma$ parametrizing the Coulomb branch is lifted. We then propose the deformation in figure \ref{pq_4} to be the one we are looking for. 

A further check for the validity of our proposal comes from computing the resulting low energy spectrum at weak coupling, namely at large $|h|$. Field theory analysis shows that the supersymmetry breaking deformation lifts the whole ${\cal N}=1$ vector multiplet of $SU(2)$ SYM but the vector field and the theory flows to pure YM in the IR. So, if our proposal is correct, not only the real scalar $\sigma$ but also the gaugini should be lifted by rotating the brane system as in figure \ref{pq_4}. To show that this is the case it is easier to work in the limit $g_s \rightarrow 0$ (this makes the analysis simpler but does not change the end result). In this regime the supersymmetric brane system depicted in figure \ref{pq_1} becomes a system of two parallel NS5-branes with two D5-branes stretched between them \cite{Aharony:1997ju}.  
The D5s extend along $(01234x)$ and the NS5s along $(01234y)$, which are separated in the $x$ direction by the finite length D5-branes.  Standard analysis of the boundary conditions of the low energy modes on the D5-branes shows that a full 5d ${\cal N}=1$ vector multiplet survives. 

\begin{figure}[h!]
	\centering
	\includegraphics[scale=0.42]{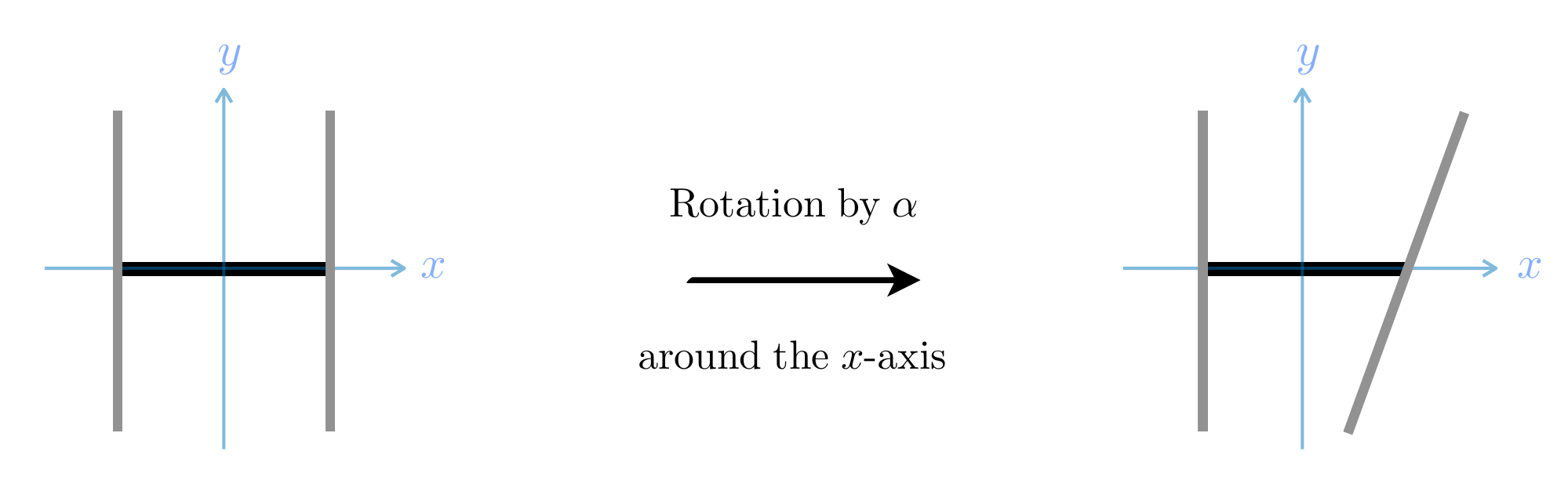}
	\caption{The supersymmetric and the supersymmetry breaking configurations at $g_s=0$ and finite $h=1/g^2$. The grey lines are NS5-branes. In the right figure the NS-brane on the extreme right is rotated around the $x$-axis on the $(y7)$ plane.}
	\label{pq_4b}
\end{figure}

In this regime, our supersymmetry breaking deformation amounts to rotate the right NS5-brane by an angle $\alpha$ around the $x$ direction, as shown in figure \ref{pq_4b}. 
The boundary conditions for a massless vector \cite{Hanany:1996ie} are still satisfied, in particular is still true that
\begin{equation}
F_{\mu x}=0~~,~~F_{\mu\nu}~\mbox{unconstrained}
\end{equation}  
where $\mu,\nu=0,\dots,4$, which are Neumann directions for both the D5 and the NS5-branes. On the contrary, the boundary condition for getting a massless gaugino are violated. In the supersymmetric configuration the following conditions are satisfied at both ends of the finite D5-branes \begin{equation}
\label{fersusy1}
\lambda_- \equiv (1-\Gamma) \lambda = 0~~,~~ \lambda_+ \equiv (1+\Gamma) \lambda~\,\mbox{unconstrained}~,
\end{equation}
where $\lambda$ is the spin 1/2 field of the D5-brane theory and $\Gamma = \Gamma_x \Gamma_7 \Gamma_8 \Gamma_9$ is a product of Dirac matrices with indices on the NS5-brane transverse directions.  The mode $\lambda_+$ is the gaugino, partner of the gauge field $A_\mu$. 
Upon rotating one NS5-brane by an arbitrary angle in, say, the $(y7)$ plane, the conditions which is satisfied at the corresponding intersection is the same as \eqref{fersusy1} but implemented by a different $\Gamma$ matrix, namely $ \Gamma' = \Gamma_x \Gamma_{7'} \Gamma_8 \Gamma_9$ where $7'$ is a transverse direction for the rotated NS5-brane, together with $x,8,9$. This mixes the $+$ and $-$ modes so that an unconstrained one, compatible with both boundary conditions at the right and left NS5-branes, does not exist anymore. Hence, the gaugino is lifted. 

The analysis of the boundary conditions at the D5/NS5 intersections provides also another way to see that after the supersymmetry breaking rotation the modulus $\sigma$ is lifted. On the D5-branes there are four scalars, $\phi^{y,7,8,9}$, associated to the possibility for the D5s to move in the directions transverse to their world-volume. The NS5-branes fixes to zero all scalars transverse to both the D5 and the NS5 branes. So, for a NS5 along $01234y$, we get 
\begin{equation}
\phi^{7,8,9}=0~.
\end{equation}
In the supersymmetric case, the D5s are suspended between two identical NS5, so $\phi^y$ survives the boundary conditions, and is nothing but the scalar field $\sigma$ in the vector multiplet of the five-dimensional gauge theory living on the finite length D5-branes. Upon rotating the right NS5-brane as in figure \ref{pq_4b} the boundary conditions there change, fixing now to zero the modes  
\begin{equation}
\phi^{7',8,9}=0~.
\end{equation}
As a consequence, imposing the two conditions at once, all scalar modes are lifted.

The origin of the instability in the rotated brane web can be seen as follows. Under a rotation by an angle $\alpha$ around the $x$-axis the $[1,1]$ and $[1,-1]$ 7-branes on which the 5-branes end get twisted by the same angle. This implies, in turn, that the two $[1,1]$ 7-branes where the $(1,1)$ strings describing the background vector multiplet live become a system of branes at angles with 2 mixed Neumann-Dirichlet boundary conditions. This provides a VEV to the D-term component of the (background) vector multiplet described by the 7-7 strings, $Y^{(ij)}_{(ab)}$. More precisely, we have 
\begin{equation}
\langle Y^{(ij)}_{(ab)} \rangle  = \hat{m}^{(ij)}\hat{v}_{(ab)}\frac{1}{\pi \alpha'} \, \tan \frac{\alpha}{2}
\end{equation}
with $\hat{m}^{(ij)}$ and $\hat{v}_{(ab)}$ being unit vectors indicating the Cartan's directions, which we choose along $(12)$ in both, as discussed around eq.~\eqref{fulldef}. Via the coupling \eqref{gauge7} this generates a mass deformation in the $E_1$ theory
\begin{equation}
\label{nonsusyc}
Y^{(ij)}_{(ab)}\, \mu^{(ab)}_{(ij)} ~ \longrightarrow \Delta {\cal L} = \langle Y^{(12)}_{(12)} \rangle \mu^{(12)}_{(12)} = \widetilde m \, \mu^{(12)}_{(12)} = \widetilde m  (|H^1_1|^2 - |H^1_2|^2)~,
\end{equation}
which is nothing but eq.~\eqref{halfdef2}.\footnote{One can reach the same result also looking directly at the 5-brane theory.} The operator $\mu_{(ij)}^{(ab)}$ is a lowest component operator of the $SU(2)_I$ current multiplet and therefore the deformation above breaks supersymmetry explicitly. 

The scalar operator $Y^{(ij)}_{(ab)}$ is in the adjoint of both $SU(2)$'s of the global symmetry group and a VEV breaks them to their Cartan generators, $SU(2)_I \times SU(2)_R \rightarrow U(1)_I \times U(1)_R$. As already noticed, this is matched geometrically: the angle between the $[1,1]$ 7-branes breaks $SU(2)_I \rightarrow U(1)_I$ and at the same time leaves only a two-dimensional plane transverse to both the $(x,y)$ plane and to the plane spanned by the two 7-branes at angles, hence breaking the R-symmetry as  $SU(2)_R \rightarrow SO(2) \simeq U(1)_R$. Note that unlike the symmetry breaking pattern on the Higgs branch we discussed previously, in this case we have a complete description of the symmetry breaking pattern, $SU(2)_I \times SU(2)_I \rightarrow U(1)_I \times U(1)_R$. This is because in this case the breaking pattern does not mix the two symmetries. Moreover, everything depends on a VEV for $Y^{(ij)}_{(ab)}$ which is a fundamental degree of freedom for the $[1,1]$ 7-branes theory.

Let us end with an important remark about the range of validity of the above brane description. In order for the brane-web after the deformation to describe a five-dimensional field theory, we need to decouple the Kaluza-Klein (KK) six-dimensional modes. Being the supersymmetry breaking scale $M_{SB} = \tilde{m} g^2$ and the KK mass $\sim \frac{1}{\Delta x}=g^2/l_s^2$, with $l_s$ the string length, the following inequality  should hold, $M_{SB} \ll g^2/l_s^2 $. So, if the supersymmetry breaking deformation is sufficiently smaller then the gauge coupling squared in string units (which is a weaker and weaker constraint the larger the gauge coupling), the brane-web still describes a five-dimensional field theory and the KK modes do not mix.

\section{Phase diagram of softly broken SYM}
\label{phdiag}

The picture emerging from our analysis is that the parameter space is divided into two qualitatively different regions. One region, which includes the weak coupling regime, where at low energy the mass deformed $E_1$ theory reduces to pure $SU(2)$ YM. A second region, symmetrically displaced around the $\widetilde m$-axis, where the global symmetry is spontaneously broken due to the condensation of the hyperscalar $H^a_i$, with a symmetry breaking pattern $U(1)_I \times U(1)_R  \rightarrow D[U(1)_I \times U(1)_R] \equiv U(1)$. This implies that a phase transition should occur between them. 

The response of the brane web in the two regimes seems to confirm this picture. For any fixed angle $\alpha$, for small enough $h$ so that an open string tachyon is generated, the $(1,1)$ semi-infinite branes are expected to recombine and separate from the $(1,-1)$ recombined brane in the transverse direction, in analogy with what happens with systems of branes at angles in flat space, see e.g. \cite{Hashimoto:2003xz}. Space separation induces a VEV for the hyperscalars, as discussed in section \ref{E1sum}, and the global symmetry is spontaneously broken. On the contrary, for large enough $h$ the tachyon disappears from the spectrum and brane recombination is disfavored (notice, further, that the larger $h$ the heavier is the $(2,0)$ 5-brane which keeps the half $(1,1)$ and $(1,-1)$ 5-branes apart, working against brane recombination). Hence in this regime the brane web does not break and the VEV of the Higgs field vanishes. 

The tachyon dynamics may suggest a second order phase transition around $|h| \sim \sqrt{\widetilde m}$ where the tachyon becomes massless. However, we do not know the full tachyon potential and present understanding of brane-web dynamics should be improved in order to gain full control on the fate of the brane system and so to be conclusive about the nature of the phase transition.

A more concrete handle on the fate of the instability may come from an effective field theory approach. Clearly, by breaking supersymmetry, the exact scalar potential is expected to contain more terms than just quadratic ones, eq.~\eqref{halfdef3}. While we cannot find its complete expression, there are two regimes where we can be quite safe about its form. 

Let us start from the $E_1$ supersymmetric fixed point and sit at a point of the Higgs branch where the hyperscalar field VEV is very large. In this regime the low energy dynamics is described by the massless hypermultiplet $\cal H$ only, since all other degrees of freedom have masses of order $\langle H \rangle$. 

Let us now switch on the non-supersymmetric mass deformation. This lifts the modulus and a potential is generated. Requiring that no singularities arise as $\widetilde m \rightarrow 0$, one can easily conclude by dimensional analysis that, besides the quadratic term, only operators with negative powers of $H$ can appear. This ensures that the absolute minima at $\langle H \rangle \rightarrow \infty$ predicted by the leading order analysis survives. Hence we conclude that on the $\widetilde m$-axis the global symmetry is indeed spontaneously broken and the potential is unbounded from below.\footnote{Depending on the specific form of the higher order corrections local minima can appear at finite distance in the moduli space, but these do not change our conclusions. In principle, it is also possible that a competitive unstable vacuum shows up at vanishing VEV for $H$. This cannot be excluded, but the brane-web description of the supersymmetry breaking deformation, figure \ref{pq_4}, suggests this not to be the case. As already emphasized, for vanishing $h$ the 5-branes tend to recombine and separate in the transverse space, hence providing a non-vanishing VEV for $H$. That the end-point of this process is a reconnected, stable non-supersymmetric brane web is very unlikely and we exclude such possibility.}

Let us now consider the other extreme regime, $\widetilde m=0, h \not =0$. In this regime the hypermultiplet is a massive BPS state at threshold and so we expect no corrections to the potential other than the supersymmetric mass term
\begin{equation}
\label{m=0}
V(H,h) = h^2 H^2~.
\end{equation}
In particular, and more relevant to us, the minimum of the potential is at $\langle H \rangle =0$ (as dictated by ${\cal N}=1$ SYM, which is in an unbroken phase). So, as anticipated, there should exist a curve in the $(h,\widetilde m)$ plane where a phase transition occurs, separating a region where the $U(1)_I \times U(1)_R$ global symmetry  is preserved from a region where it is spontaneously broken. 

\begin{figure}[h!]
	\centering
	\includegraphics[scale=0.45]{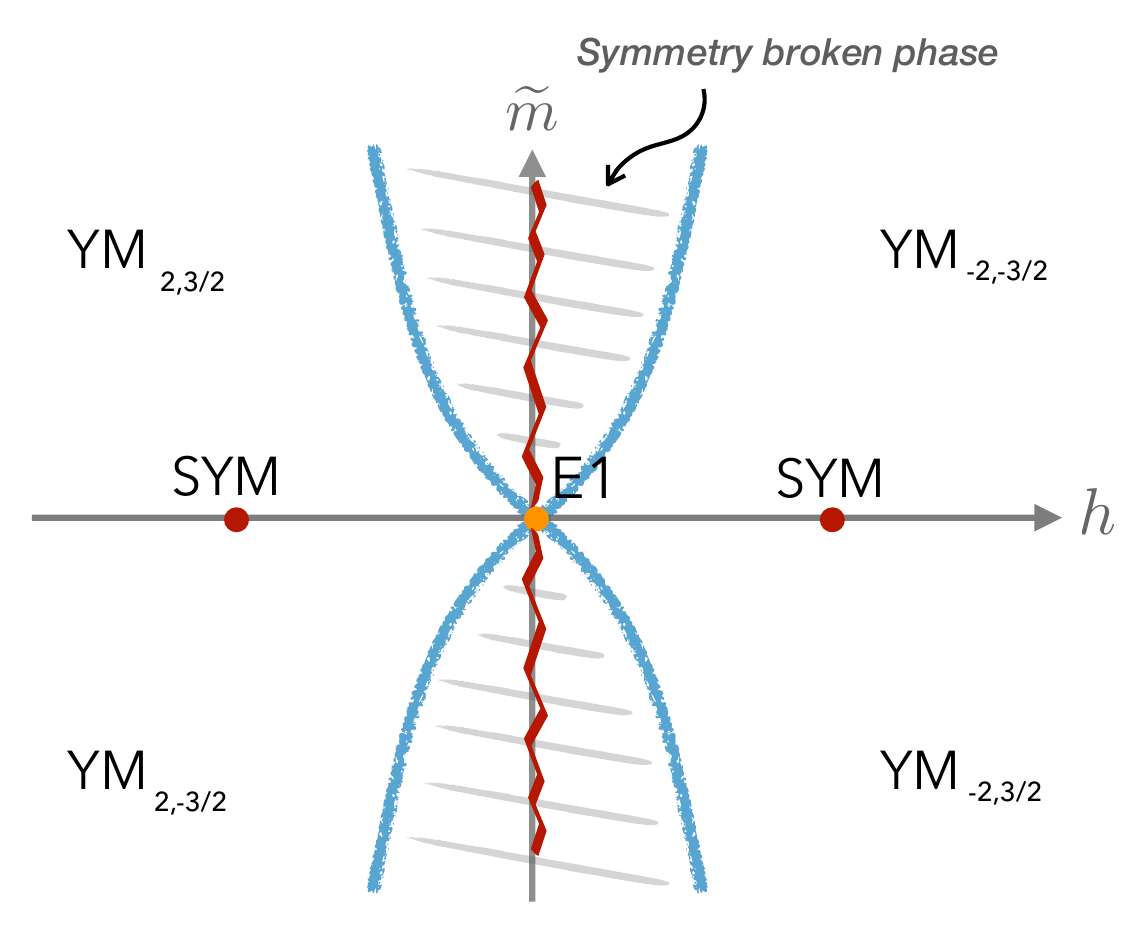}
	\caption{Phase diagram of softly broken $E_1$ SCFT. The four regions described by pure $SU(2)$ YM enjoy different $U(1)_I$ and $U(1)_R$ background CS levels, as indicated. The instability region (red wavy line) can be confined on the $\widetilde m$-axis or extend in part or all the symmetry broken phase, depending on the exact form of the potential. Along the blue thick lines a phase transition occurs, which can be first or second order. In the latter case, note that the critical line does not represent a non-supersymmetric one-dimensional conformal manifold. The deformation out of the $E_1$ fixed point is a relevant one, which is seen as irrelevant from the IR theory point of view. The ratio $h^2/\widetilde m$ is a marginal parameter but its value is tuned on the critical line. Hence, if a CFT actually exists, it is unique. 
}
	\label{ps_1}
\end{figure}

For both $\widetilde m$ and $h$ non-vanishing, the potential should interpolate between the above regimes. Computing its exact expression is a daunting task (in brane-web language this would correspond to compute the full tachyon potential). Moreover, as soon as $h \not =0$ it is not at all guaranteed that the only light field is the hypermultiplet ${\cal H}$. So, in principle, the potential could depend on more scalar fields than just the hyperscalars.  
Still, even under the restrictive assumption of one light field only, some lessons can be learned, just using continuity arguments and consistency of various limits. In particular, one can show that as soon as $h \not =0$ the instability may disappear and global minima at finite distance in field space may arise, corresponding to a stable symmetry broken phase out of the $\widetilde m$-axis. On the other hand, without further insights one cannot be conclusive about the order of the phase transition separating the symmetry broken and the symmetry preserving phases. This depends on the exact form of the potential, which we do not know. A qualitative picture of the phase diagram is reported in figure \ref{ps_1}. 

If the phase transition is second order and the corresponding CFT an interacting one, this could be viewed as a UV-completion of pure YM $SU(2)$.
Upon integrating out the massive gaugini and the real scalar at one loop, the effective YM coupling reads   
\begin{equation}
1/g^2_{YM} = 1/g^2 - ag^2 |\widetilde m|~,
\end{equation}
with $a$ positive and $1/g^2=h$. This suggests that the theory becomes strongly coupled  at finite $h$, where the YM coupling diverges. Past infinite coupling the theory enters a different phase where one could expect, on general ground, an instantonic operator to condense. This nicely agrees with the phase diagram described in figure \ref{ps_1}, which was obtained looking at the Higgs branch dynamics and which shows that in the confining phase an instantonic operator do condense, in fact. 

Different approaches are been pursued to look for non-supersymmetric interacting fixed points in five dimensions. In particular, as far as UV-completions of  $SU(2)$ YM are concerned, recent progress using lattice simulations and the $\epsilon$-expansion can be found in  \cite{Florio:2021uoz} and \cite{DeCesare:2021pfb}, respectively. We think that intertwining all these approaches, including brane-web techniques discussed here, could be very promising to make progress in this quest.

In this paper we have considered one specific example within the class of rank-one superconformal field theories discovered in \cite{Seiberg:1996bd} (and higher rank versions also exist, see e.g. \cite{Intriligator:1997pq} - and \cite{Bergman:2020myx} for a brane-web description). One could wonder whether supersymmetry breaking deformations analogous to the one considered here could actually lead to different scenarios, where the global symmetry is not broken at strong coupling and no instabilities arise. This would give more chances for an actual CFT to emerge at strong coupling. The role that the hyperscalar parametrizing the Higgs branch played in the present model, both in (spontaneously) breaking the global symmetry and giving an instability, may suggest to look for models where there is no emergent Higgs branch at infinite gauge coupling (a similar approach has been pursued for theories in four dimensions in, e.g., \cite{Cordova:2018acb}). In this respect the $E_0$ and $\widetilde E_1$ theories \cite{Morrison:1996xf} can be promising set-ups to be considered. Work is in progress in this direction.

\subsection*{Acknowledgements}

We thank Thomas Dumitrescu and Pierluigi Niro for collaboration at the early stage of this work. We also thank Riccardo Argurio, Francesco Benini, Stefano Cremonesi, Lorenzo Di Pietro, Thomas Dumitrescu, Carlo Maccaferri, Pierluigi Niro and Diego Rodriguez-Gomez for discussions. Finally, we are grateful to Lorenzo Di Pietro for very useful feedbacks on a first draft version. This work is partially supported by INFN Iniziativa Specifica ST\&FI.

\end{document}